\documentclass[a4paper,14pt]{article}

\begin{document}

\qquad FUNDAMENTAL  FIELD THEORY IN TEN DIMENSIONS

\qquad\qquad\qquad AND THE EARLY UNIVERSE

  \qquad\qquad \qquad \qquad\qquad         Ajay Patwardhan

\qquad Physics Department, St Xaviers college, Mumbai, India

             \qquad \qquad \qquad\qquad\qquad\qquad   ABSTRACT

A unified  field theory in ten dimensions, of all interactions, can describe the high energy processes occurring in the early universe. In such a theory transitions that give properties of the universe can occur due to the presence of algebraic and geometric structures.

A correspondence between  theory and observations of the universe is made, to obtain a new interpretation and properties. This paper consists of a  field theory and cosmological model of dark and normal energy and matter, cosmological constant, acceleration and inflation in the early universe.

\qquad \qquad \qquad\qquad\qquad I. INTRODUCTION

In previous papers[1],$ (1)$ the consistent occurrence of commutative and noncommutative geometry structure, and quantum and symmetry group structures was shown, $(2)$ the unified gauge theory for all the interactions and the holonomy based reduced dimension Calabi Yau spaces were defined,$ (3)$ the spectrum of operators in spaces with genus studied,$ (4)$ the presence of topological solutions and transitions were described with partition functions.

In this paper the new properties obtained in the previous work are placed in the context of the early universe. It is shown that there is a possibility of obtaining a new insight about three and perhaps more of the properties of the early universe.$ (1)$ The inflation phase which leads to the general relativistic and Newtonian cosmological models.$ (2)$ The primordial black holes.$ (3)$ The creation of normal and dark matter and energy. In this paper the available structures in the fundamental field theory are used to obtain these properties.

The topologically nontrivial solutions for the Calabi Yau spaces occur at scales when commutative geometry emerges from non commutative geometry. Euler number based equivalence classes are used to describe them. As symmetry breaking gives each gauge field its independence, normal and dark matter are produced. The quantum group has special cases of Fermionic and Bosonic kind which lead to normal matter. The other solutions for the quantum group are possibly dark matter.[2,3,4]

 The spectrum of the differential operators, evaluated in the spaces with genus, and in genus changing transitions, implies that the vacuum energy released for the gauge fields contributes to the normal and dark energy. The topological solutions with the holonomy based quantum field theory can create primordial black holes and  geons , when the gravity gauge field is extracted from the unified gauge field theory. Path integral based calculation of these processes gives some results in simple models.[1]

 The inflation phase in the early universe is seen as the dynamics of the particular 6 D Calabi  Yau space $\otimes$ 4 D Riemannian space on which the unified gauge field theory acts. In this manner there is a possibility of understanding some properties of the early universe in terms of the structures available in a fundamental field theory.

The recent observational evidence for cosmological parameters is used to obtain a standard cosmological model, in which the ten dimensional theory gives a new interpretation of the dynamics.[2,18]

\qquad II DESCRIPTION OF THE $10$ DIMENSIONAL FIELD THEORY AND UNIVERSE MODEL

In previous work[1] I had shown that topological index (Euler characteristic) in Partition functions has applications in Field theory and Statistical Mechanics transformations.$ 10 $ dimensional spaces with compactification of $ 6 $ dimensional Calabi Yau spaces and $ 4 $ dimensional spacetime is a fundamental description of nature in high energy physics and in the universe.

 This paper develops an approach to obtain a cosmological model within this framework and within observed parameters. The early universe during inflation is considered to be a Calabi Yau space of six dimensions$\otimes$Riemann space of four dimensions. At late times the six dimensional space is compactified to extremely small dimensions and attatched locally to the spacetime which is globally Riemannian and locally Lorentzian.

The compactification of higher dimensional spaces in the early universe  produces a $ 4 $ dimensional Riemannian geometry  spacetime with an expanding $ 3 $ space in inflation and post inflation epochs. The expansion of the  $ 3 $ surface which becomes  the observable universe and the collapse of the $ 6 $ dimensional Calabi Yau spaces gives rise to topology and geometry tranformations.

 The resulting contributions to vaccuum and Casimir energies of the fields to the inflation and post inflation expansion and the creation of normal and dark matter and energy give a cosmological model. In the energy range $10^{18}$ GeV to $10^{12}$ GeV, the symmetry breaking successively occurs for gravity to go from strong to weak coupling constant$ 10^{-2}$ to $10^-{34}$ and for supersymmetry to break weakly from $!0^{16}$ GeV to $ 10^{12} $ GeV, and for the ten dimensional theory to go to the Calabi Yau $(6)\otimes$ Riemann $(4)$ space as given in SUSY/Supergravity theories. The scalar, fermionic matter and bosonic gauge fields in Einsteinian spacetime occur.

An analog could be obtained as a sponge or honeycomb that is squeezed so that the contents come to the boundary surface from the bulk interior , which collapses its vesicle structures and changes topology. Spacetime foam models and spin polymer network models in quantum gravity, geometry, cosmology are an alternative at an earlier stage in the universe. The  $ 3 $ space of the universe as a $ 3 $  brane with scalar and gauge fields on the compactified $ 5 $ brane also lead to models in cosmology.[9,10]

In the Partition function the weightage is higher for simpler topologies, as these have smaller genus numbers and hence less negative Euler Numbers ( which are in the exponent); and these topologies are preferred during quantum fluctuations and statistical averaging over Calabi Yau space compactification.

 In early cosmology the energy scales: Planck, string, SUSY breaking and GUT energy scales create possible transition points at which the  $ 10 $ dimensional space essentially becomes a $ 3 $ space expanding in time , with extremely microscopic $ ( 10^{-18} $ meters)scale higher dimensional spaces attatched locally . So the'' boundary''(subspace) expands at the expense of the'' bulk''(subspace)interior; thus creating the observable universe with its contents and dynamics.

Non commutative geometry and fermionic and bosonic coordinates were considered with Bogoliubov transforms. These can be found for the expanding geometry. The Supersymmetry conditions in an expanding geometry , included in the Unified field theory give a constraint on the collapsing Calabi Yau spaces and the breaking of Supersymmetry.

 The spectrum of operators on topologies, given by Fundamental group representations and geodesic length spectrum, gives results to obtain the Casimir and vaccuum energies. A holonomy representation for the fields was taken to enable expectation values of field variable products to be found. These definitions from the previous work are useful in this model.

Observational evidence has increased in parametrising the observable universe. Recent reviews state  a rapid early stage and a slow late stage of accelerated expansion. WMAP results have given anisotropy and polarisation of microwave background and given support for the inflation stage described by a $- \phi^2 + \phi^4 $ term scalar field.

 Dark energy and matter estimates and equations of state contributing to the cosmological model specify limits on $ \Omega, \rho, q, H $ for the relative contributions from the normal and dark energy and matter. Beyond $ z = 5 $ the variables appear to tend to fixed values, suggesting a asymptotically flat universe.

The polarisation plots indicate that a vector field ( which is not the gradient of a scalar field), with non zero curl, exists globally; or there is non zero vorticity in the universe observed $ 300,000 $ years ($10^13 $ seconds) after the big bang. The assumption that only  the scalar non linear field drives inflation ;and the additional gauge fields and bosonic and fermionic fields arise later is questionable.[2,3]

 There is a remnant of the supersymmetric Fermionic/Bosonic field ( which could be written in Superspace notation in a common framework ( with commuting and anticommuting number systems), which evolves into the standard model fields after Super symmetry breaking energies/ temperatures occured($10^{25} $ eV and $ 10^{-12} $ seconds. In the $10^{-30} $ to $ 10^{2} $ seconds --- $ 10^{12} $ eV energy for the inflation to complete and particle creation to occur there would have to be more inputs into the physics from this unified field. In the standard model of particle creation a non equilibrium phase sets in post inflation stage creating particle/antiparticle asymmetry.

\qquad III. THE FUNDAMENTAL SUPERSYMMETRIC UNIFIED FIELD THEORY MODEL

The unified field theory and generalised field theory developed in previous papers gives a common representation for ``q'' commutators of which supersymmetric Fermi/Bose systems are special cases and Bogoliubov transforms consistent with gauge and (non)commutative structures were defined. In an accelerating geometry these transforms can be written as follows.[4]

Functions $ cosh(g(k,a))$ and $  sinh(g(k,a))$ for bosonic, and $ cos(f(k,a))$ and $ sin(f(k,a)) $ for fermionic systems are used in the transforms of the operators $\hat{b}(k)$ and $\hat{b}^{\dag}(k)$ for bosonic, and $\hat{c}(k) $ and $\hat{c}^{\dag}(k)$ for fermionic variables respectively. $k$ is the momentum/wavevector and $a$ is the acceleration. 

The functions $g(k.a) $ and $ f(k,a) $ are functions of the acceleration and depend on the expanding geometry model. For some standard cases these are given for quantum field theory in curved spacetime and are increasing functions of acceleration; actually dependent on the scale factor $s(t)$ in the cosmological model as given below.

For the bosonic system the transform is 

\begin{displaymath}
\left(\begin{array}{c}b^{'}(k)\\b^{'}(-k)^{\dag} \end{array}\right) = 
\left(\begin{array}{cc}cosh(g(k,a)) & sinh(g(k,a))\\ sinh(g(k,a))& cosh(g(k,a))\end{array}\right)\left(\begin{array}{c}b(k)\\b(-k)^{\dag}\end{array}\right)\end{displaymath}

Similar expressions for the fermionic system are

\begin{displaymath}
\left(\begin{array}{c}c^{'}(k)\\c^{'}(-k)^{\dag} \end{array}\right) = 
\left(\begin{array}{cc}cos(g(k,a)) & sin(g(k,a))\\- sin(g(k,a))& cos(g(k,a))\end{array}\right)\left(\begin{array}{c}c(k)\\c(-k)^{\dag}\end{array}\right)\end{displaymath}

The Number operator has vaccuum expectation values

 $ sinh^{2}(g(k,a)) $ and $ sin^{2}(f(k,a)) $ 

These give energy  and density conditions. In uniformly accelerated frame this results in a typical thermal like  bosonic and fermionic distribution with acceleration playing the role of the temperature parameter. The values of these vaccuum expectation values are very large when acceleration is large in inflation , and become very small in late universe when acceleration is small. They are highest for scalar bosonic fields and that gives the major contribution to dark energy and matter during inflation , with the vaccuum expectation value occuring as the cosmological constant.

The presence of a vector gauge field $ A ^{(n)}$ that can have additional degrees of freedom ; such as in SU(N) gauge groups, allows a Holonomy representation and expectation values of products of field variables written in terms of the partition function,in the multiparticle states can be found.[5,6,11]

Holonomy : \qquad $Tr P e^{ \oint A^{n}T_{(n)}dx}$

Partition functional: \qquad $\int d[A]e^{(-S[A])}$

Vaccum expectation value: $ < vac|AAA--|vac>$

 Similar expressions for all the fermionic and bosonic fields are defined.

 Non zero values are expected in the accelerating geometry and the detailed form of the functions and their contributions to the number and energy density will be model dependent. From these processes, Supersymmetric forms of dark matter (such as neutralinos) and normal matter are created and their compositions and quantity are obtained. SUSY breaking occurs before GUT scale in the inflationary early universe.

 The supersymmetric transforms can be applied to the Calabi Yau spaces. Since the six dimensional spaces allow three complex coordinates,and a $CP^{3}\otimes CP^{3}$ representation . For spaces with Euler number $-6 $ as the simplest topologies which are preferred in compactification, cyclic coordinates could be chosen
 $ \prod{ \zeta_{i}}$, for example on a complex three sphere and  torus; or six real coordinates can also be used. These spaces are called Tian Yau spaces, and these give the best fit to the standard model of particle physics at below GUT scale. [4,11,12]

Using $ (X^{\mu}, \Theta, \bar{\Theta}) $ as the Superspace coordinates; that is $ 4+3+3 $ coordinates, the bosonic $X^{\mu}$ and the fermionic $ \Theta, \bar{\Theta}$ transforms are defined that are supersymmetry transformations.

$[X_{\mu}X_{\lambda} - X_{\lambda}X_{\mu}] = \delta_{\mu\lambda}L_{ss}^{2}$

The $L_{ss}$ is a supersymmetry length (energy) scale. SUSY is exact for higher energies. Above the  GUT energy scale the $L$ can be taken as non zero and then set to zero for broken supersymmetry. The coordinates $X^{\mu}$ form a $4$ dimensional Riemannian geometry spacetime and become commuting.

$ \Theta_{\alpha} \Theta_{\beta} + \Theta_{\beta}\Theta_{\alpha}= \delta_{\alpha\beta}$

And similarly for $\bar{\Theta}$

These coordinates for the six dimensional Calabi Yau manifolds describe the Internal symmetry spaces and at GUT scale , the compactified spaces are expected to give the standard MSSYM model.

Supersymmetric transforms preserving these relations are defined by introducing supergaugefields $\sigma$. Using these the covariant derivatives can be defined and the superfields calculated .

 Hermitian quadratic forms for the resulting supersymmetric fields are used to obtain the Action and hence the partition function. Evaluating these typically leads to a gaussian multiplied by a prefactor which depends on the dimension and choice of compactifed manifold. The topological collapse of the six dimensional SUSY CY spaces is expected to provide the matter and energy and the expansion of the four dimensional spacetime during inflation.

$ \Theta^{'}_{\alpha} = \Theta_{\alpha} - i \sigma^{\mu}_{\alpha\alpha^{'}}\frac
{\partial}{\partial X^{\mu}}$ .

And similarly for the adjoint $\bar{\Theta}$

The derivative operators are

$D_{\alpha} = \frac{\partial}{\partial\Theta^{\alpha}} + 2 i \sigma^{\mu}_{\alpha\dot{\alpha}} \bar{\Theta}^{\dot{\alpha}}\frac{\partial}{\partial X^{\mu}}$

The compactification over possible geometries and topologies is obtained from the partition function. Keeping the simplest topologies as the end result of the collapse; the effective energy density of the vaccuum can be found using the energy spectrum from the geodesic lengths and the Euler characteristic. The largest geodesic length in the domain gives the smallest energy, and the smallest negative number Euler characteristic dominates the summation over all configurations.[1]

 Thus it is sufficient to consider during the collapse the simplest topologies and find therein the maximum geodesic length, to obtain a bound on the vaccuum, and Casimir energy. These geodesic lengths are simply related to the size of the domains for simple geometries, and the expanding spaces and collapsing domains, imply a variation in the energy spectrum. The minimum and maximum length geodesics determine the bounds on the energy spectrum, energy goes as inverse square of geodesic length.

The Casimir energy for simple cases, sphere, cylinder and torus is calculated for scalar and vector fields in a number of dimensions. The energy scales as $ E =\frac{ C_{g}}{d^{\alpha}}$ ; where  $ C_{g} $ is a geometry and dimension dependent positive or negative number, and $d$ is the maximum separation between boundaries of the domain, with the exponent $\alpha$ also dependent on the geometry and dimension of the space and a  positive number. Contributions from all the fields present scalar, vector, tensor and spinor need to to be evaluated to get exact expressions.[7]The averaged Casimir energy from all the domains is needed.

 For example with a vector gauge field, the Casimir energy scales as $ -0.1/d $ in a sphere of radius $d$, $+ 0.006/d^{2}$ in a cylinder of radius $d$, $-1/d^{3}$ between planes separated by distance $d$ , all in three dimensions. For a space $ M^{4}\otimes S^{N}$ with $N=6$ dimensions, it is $-10^{-5}/d^{4}$. The value is positive for spinors and negative for scalar fields. Thus for compactified scales of $10^{-18}$ meters or smaller this can become a large number. The sum of all contributions is estimated at the SUSY breaking inflation to give a figure $>10^{20}$eV. This makes the cosmological term large during the collapsing CY spaces and inflating spacetime epoch of the early universe.[20,21]

 The large energy density created in a small region will not cause horizon problems by collapse as during inflation quantum fluctuations dominate. However it may create primordial black holes in the four dimensional spacetime that have a quantum evaporation. It is geometric and topological energy that is being released , as potential energy converts to energy available for rest energy and kinetic energy of matter and radiation like energy.

This Casimir process for the six dimensional subspaces ,  along with the partition functional expectation values for matter energy fields in four dimensional spacetimes creates respectively the cosmological constant term and the right hand side of the Einstein equation, in the early universe. At late times the classical stress energy tensor occurs on the right hand side , while the cosmological term remains as it is. The identification of the scalar field value with the cosmological term in inflation and the dark energy with the cosmological term in the late times is often done. The mechanism to produce these is explained in the model in this paper.

The vaccuum energy released due to the collapse of  $ 6 $ dimensional Calabi Yau spaces contributes to the normal and dark energies in the $ 4 $ dimensional spacetime. This contribution also drives the inflation and expansion of the universe. Formally a $10$ dimensional theory written in superspace would allow superfields and action functionals to be defined.

 The collapse or compactification of the $6$ dimensional subspaces allows those dimensions to be integrated out in the functional average; thus contributing a prefactor to the remaining four dimensional integral. This in turn contributes to the dynamics of the four dimensional spacetime.

$\sum dim(M)^{\chi(M)}\int D[\phi,\psi,A]e^{(-S[\phi,\psi,A])}<\prod [ \phi\psi A]> $

 gives the expectation value of field operators. The Action could be written as a Hermitian quadratic form

$S = \int d^{D}(X,\Theta,\bar{\Theta})$Tr$(g^{2}_{1}\phi^{\dag} F_{1}\phi + g^{2}_{2}\psi^{\dag}F_{2}\psi + g^{2}_{3}F_{3}\wedge *F_{3})$

for the scalar, spinor and gauge fields; $\phi,\psi,F_{3}$ , with their coupling constants $g_{1},g_{2},g_{3}$ respectively. $F_{1}, F_{2}$ are the multiplet coupling matrix derivative operators, for minimal gauge covariant coupling acting on the fields and their adjoints.

What is the value of dark energy density and of  the cosmological constant during inflation and post inflation as compared to its expected values in late time universe. Observed evidence of the latter suggests tiny acceleration and the former requires large accelerations from theory. In the inflation case the acceleration is a parameter in the Bogoliubov transforms and hence a consistency argument can be made for its value.

 In late time the physics is governed by remnants modifying the standard model and accurate observations can help to fix the parameters of the remnants. From both these initial and final conditions an interpolation to the times of $ 10^{2} $ seconds to $10^{12} $ seconds can be made at which time the observations have become reliably available.

 IV.COSMOLOGICAL MODEL: CALABI YAU$(6)\otimes$RIEMANN$(4)$SPACE

The present understanding of cosmological theory and observations is summarised as required for this paper. The fractional composition is dark energy $0.65$ , dark matter $0.30$, baryonic matter $0.04$, radiation $ 0.002$. Data for Doppler shift $z = \frac{\dot{ s(t)}}{s(t)} -1$, where $s(t)$ is the scale factor in the metric, is known.

 It shows for $0<z<1$ indicates acceleration, for $1<z<10$ decceleration, and for $z>10$ acceleration. Rapid acceleration in the inflation stage.

 Typically pressure $  P<0 $ for acceleration and $ P>0$ for decceleration.The equation of state $P=w \rho$ for stiff matter is $w=1$, for radiation $w= \frac{1}{3}$, for dust $w=0$, for dark matter and energy there are bounds known; and $w=-1$ is best value.

 Dark energy is creating the present stage of acceleration, that began when the universe was $0.60$ of its present size.

$P=-\rho=-\frac{\Lambda}{8\pi G}$, related to cosmological constant $\Lambda$.

satisfies the condition $\rho + 3P < 0$, in the equation for acceleration, as given below. 

The origin of dark matter and energy is an open question , although the best answer in the present paper is to take dark matter as neutralino, and photino as the surviving lowest mass supersymmetric partners created at strong SUSY breaking scales; which could be upwards of $10^{4}$ GeV. The dark energy is occuring as the cosmological constant term , but it could have been created due to higher dimensional effects and scalar fields.

 Consider the compactified Calabi Yau spaces as described in the previous section and now consider the cosmological model for the Riemannian spacetime. During the inflation epoch $10^{18}$ GeV to $10^{12}$ GeV , the processes described earlier occur. They lead to a new interpretation of the revised ``standard ``model for inflation and post inflation cosmology.

If all observers in the universe were to agree on a single model of the universe, from their local observations of the observable universe; then the  geodesic coordinate system may be preferred, as it is `` locally consistent'' with the expanding geometry. In these coordinates the Ricci and Riemann tensors have a simpler form.

In the standard model for cosmology, viz. Friedmann, Robertson, Walker, De Sitter models; the Einsteinian and Newtonian methods to derive dynamical equations of the cosmologies can be used to write the equations governing the $R=R_{0}s(t)$, where $s(t)= e^{Ht}$ is the scale factor of the $3$ sphere, occuring in the metric of the four dimensional spacetime

 $dS^{2}= dt^{2}-s^{2}(t)(dr^{2}+r^{2}d\Omega^{2})$:[15]

$ 2(\frac{\ddot{R}}{R}) + ({\frac{\dot{R}}{R}})^{2} + \frac{k}{R^{2}}$  = $ \Lambda - 8 \pi G \rho $ ; $ c = 1$.

$({\frac{\dot{R}}{R}})^{2} + {\frac{k}{R^{2}}} = {\frac{8 \pi G \rho}{3}}+{\frac{\Lambda}{3}}$

With ${\frac{\dot{R}}{R}} = H $ is the  expansion parameter (Hubble) and

$ q = -{\frac{R \ddot{R}}{\dot{R}^{2}}} $ is the acceleration parameter. Recall this gives value of acceleration to be used in the Bogoliubov transforms, in previous section. 

The  $ {\frac{\ddot{R}}{R}} = {\frac{\Lambda}{3}} - {\frac{4 \pi G ( \rho + 3 P )}{3}}$ ; gives conditions for acceleration positive or negative.

It is modified to include fractional $n_{i}$ contributions of each form of matter and energy, normal and dark, with their equation of state $P=w \rho$;

$ {\frac{\ddot{R}}{R}} = {\frac{\Lambda}{3}} - {\frac{4 \pi G \sum_{i} n_{i} ( \rho_{i} + w_{i} \rho_{i} )}{3}}$

The radiation and matter (density and pressure) terms scale as inverse fourth and third power of $R$ respectively. $P =\rho/3 $ for radiation and $P=-\rho$ for dark energy.

The cosmological constant $\Lambda$ occurs as a $\Lambda g_{\mu\nu}$ term in the Einstein equation

$R_{\mu\nu} - \frac{R g_{\mu\nu}}{2} +\Lambda g_{\mu\nu} = - \frac{8 \pi G T_{\mu\nu}}{c^{4}}$

This occurs due to a term in the Action

 $ S = \int d^{4}x\sqrt{g}(R+\Lambda)+ \int d^{4}x\sqrt{g}L $

The local Lorentz invariance required in the theory requires that the expectation value of the stress energy tensor $ T_{\mu\nu}$ should be a parameter times the metric tensor: $ \Lambda g_{\mu\nu} = 8 \pi G <T_{\mu\nu}>_{vac} $ and $ <T_{\mu\nu}>_{vac}$ = diag$(\Lambda,-\Lambda,-\Lambda,-\Lambda)$; hence of the same form as the cosmological term in Einstein's equation.

 Each type of spin $j$  particle contributes $(-1)^{2j}(2j+1)$ prefactor to the integral. These integrals scale with $M^{4}$ for an appropriate mass scale; Planck, SUSY, GUT or some other mass scale. To obtain consistent numbers has been difficult since the composition of the universe has contents , dark matter and energy, whose observed parameters are not enough to make a fundamental theory for them.

 We could therefore take the origin of the cosmological term as the vaccuum expectation value of the matter and energy tensor. The classical stress energy tensor is retained as source term in late times, and dropped during the inflation epoch.

 In the framework of this paper, in the rapidly accelerated geometry of the inflationary universe, the collapsing six dimensional spaces release vaccuum and Casimir energy into the expanding four dimensional spacetime. This contributes to the large value of the cosmological constant in the inflation stage.

 In the late stage of expansion the contribution from the Bogoliubov transformed vaccuum drops significantly as the acceleration is very small and so the value of the cosmological constant in present epoch is very small.

Observational evidence indicates that acceleration parameter $-q$ is slightly increasing that is the ${\frac{d}{dt}}(\ddot{R})$ is positive. In early universe during inflation this grew rapidly. For asymptotically flat universe $k$ is nearly zero, and this is consistent with data. The cosmological constant $ \Lambda $ , actually had different values in inflation than it has now and it indicates that different mechanisms contribute to these different values.

 The $ \rho + 3 P $ term is actually dependent on the equation of state and so for each kind of normal and dark energy and matter this should be separately taken. Generally it should be stated as  $ P= w \rho$, and so   $ \rho + 3 w \rho $; with values for $ w $ fixed from theory and observation. Newton's Gravitational constant is taken to have reached the observed value , even though strong gravity regime could exist from Planck scale to string scale.

If the total $ \rho + 3 w \rho $ is set to zero; then the equation is solved to get the

  $ R(t) = R_0 e^{(\sqrt{{\frac{ \Lambda}{3}}} t)} $ ,

 This explains an exponential growth rate during inflation, but with a large $\Lambda$. This growth is usually explained using the scalar field vaccuum expectation value. $w < -1/3 $ is needed to accelerate the expansion. To close the set of equations it is necessary to evolve the density.

For total ${ \dot{\rho}} = -3( {\frac{\dot{R}}{R}})(\sum_{i} n_{i} (\rho_{i} + w_{i}\rho_{i})) $;

Fractional $n_{i}$ contributions  should be added from each component of dark and normal matter and energy. From observations $n_{i} = 0.7,0.25,0.05$ respectively for dark energy, dark matter and normal matter/energy; along with the appropriate $w$ inserted for the equation of state.$ \rho_{\Lambda} =2 \rho_{M}$ is the present time value for $z=0$.

 However for $R<<R_{0}, \rho_{m} >> \rho_{\Lambda}$ and vice versa. The turn off happens at $\ddot{R} = R_{0} \ddot{ s(t)}=0$.

 For $P>0$ decceleration and for $ P<-\frac{\rho}{3}$ acceleration gives $\frac{P}{\rho} < \frac{-1}{3}$ for dark energy.

 The zero acceleration condition from the

$\frac{\ddot{s}}{s} = \frac{-4 \pi G \rho_{0}}{3 s^{3}} + \frac{\Lambda}{3} =0$ 

condition gives the value $\rho_{\Lambda} = \frac{\Lambda}{ 8 \pi G}$

as $s^{3} = \frac{\Omega_{m}}{2 \Omega_{\Lambda}} =\frac{0.3}{2 \times 0.7}$

 If the cosmological term is understood to originate as explained in the previous section, then the large $R$ limit allows us to drop the inverse power terms in $R$ and again have the asymptotic exponential expansion with a weak cosmological term in the exponent; also interpreted as (slow) acceleration.

 The origin and constancy of the cosmological term have raised many questions. The way it occurs in the cosmological theories can be made consistent with the observed values of parameters only with some new fundamental theory; that is outside general relativity, but of which general relativity must be a part. That is attempted in this paper.

For the density this gives a decreasing function and the critical density $ \rho_c $ can be used to find the ratio $ {\frac{ \rho}{\rho_c}} = \Omega $. Theory and observation appear to be consistent for $\Omega = 1$ and a spatially flat universe.

 For correlating with observations the quantities are expressed as functions of the Doppler shift factor $z$; however beyond $5<z<100$ they tend to their fixed values, such as acceleration $q=1/2$, and $w=-1$ or $P=-\rho$ for dark energy. The derivation of the equation of state for dark energy and matter and the Free energy produced in inflationary expansion remains an open question, although some models for the action and partition function have been reported.

It is useful for parametrisation of the $R(t) = R_{0} s(t)$ with a  trial function $s(t)= u t^{v} e^{(\gamma t)}$, where $u,v, \gamma $ are parameters that can be related to observed data, which is usually given in terms of the shift $z$, in the early and late times universe. Simulations using this form for $s(t)$ work better in the asymptotic range, but they are indicative rather than definitive, about the trends and the parametrisation.

 As stated in the most recent post  WMAP- 3 results analysis [9,18] the correct model for the universe is turning out to be simpler than expected and a revised ``standard'' model is expected. The pre inflation quantum gravity, quantum cosmology and string/brane cosmology regime is an area of active theoretical research that has to give the correct match to the initial conditions for the inflation models.

 These inflation models have now been revised to produce the good fit to the observed data at the $300,000 $ years or $10^{12}$ seconds time, which is $ z = 1000$. But a theoretical basis for the intermediate time of $ 10^{-2} $ seconds to $10^{10}$ seconds requires detailed work , as there is no data obtainable from this epoch by direct observation [2,10]. In this regime the particle/antiparticle asymmetry, the quark gluon condensation and other phase transitions are expected and the cosmological model is not expected to be simple. 

 The subsequent content, dynamics and model of the universe uptill the late time $10^{16}$ seconds, with a recently  observed slight increase in acceleration attributed to dark energy effects, as accurate observations from a variety of indicators are compliled. is also leading to a relatively simple form of the $\Lambda CDM --FRWD $ model. This paper provided some new interpretations of the observed effects based on the ten dimensional field theory developed in my earlier work[1].

\qquad \qquad\quad  V. INTERPRETATION AND CONCLUSION

The model presented in this paper is based on my previous work and attempts to make a simple analysis of a complicated phenomenon. The main points are:

1.There is a $10$ dimensional fundamental theory in which a six dimensional subspace compactifies as the four dimensional subspace (three surface) expands.

2.All necessary fields exist as multiplets, scalar, vector, tensor, spinor in a supersymmetric form , and the fermionic and bosonic variables can be obtained as inflation occurs.

3.The Casimir and vaccuum energy of the collapsing Tian- Calabi Yau spaces is driving creation of all forms of matter and energy and expansion of spacetime.

4.The cosmological (constant) term arises as the vaccuum expectation value of the stress energy tensor of all matter and energy fields; its value is large in inflation; as the accelerated expanding geometry gives a large Bogoliubov coeffecient. Exponential expansion with $t = 10^{-n}$ seconds, $n$ positive results.

5.The early universe standard (FRWD)cosmological model has the classical stress energy tensor of all components of normal and dark matter and energy as source for the expanding three surface of the observable universe.
6 This model is consistent with requirements and  the  results of $\Lambda CDM $ model and it is consistent with WMAP and other observations; particularly for $z>5$ the critical parameters $\Omega,q,w$ reach their fixed values as expected in a spatially flat universe.[9,16,18]

7.The late time universe has the acceleration dependent form with a small cosmological constant value and the slow exponential expansion has a $t=10^{n}$secondsfactor for $n$ positive$(12$ to $17)$.

8.A consistent theory of the observable universe and its contents and dynamics ,requires the $10$ dimensional fundamental field theory; reducing it to a four dimensional spacetime theory, in which general relativity applies.

 Considering the apparent simplicity of this model consistent with data; one recalls Einstein's remark on the comprehensibility of the universe.

\qquad\qquad\quad\qquad\qquad VI. ACKNOWLEDGEMENTS

I wish to thank the Director and the Institute of Mathematical Sciences, Chennai for the facilities and support for my visit. Discussions with Prof Sharatchandra, Prof Ghanashyam Date, Prof Romesh Kaul, and Prof N. D. Hari Dass are thankfully acknowledged.

\qquad \qquad\qquad \qquad\qquad VII.REFERENCES

[1]Ajay Patwardhan,

  (a) Unified gauge field theory and topological transitions, 
   www.arxiv.org/hep-th/0406049;

   (b)Statisticl Mechanics of Dynamical Systems and Topological

   Phase Transitions,  cond-mat/0511231;

  (c) Quantisation on General spaces, quant-ph/0211039 ;

   (d)Particle and Field Symmetries and Non Commutative 
   Geometry, quant-ph/0305150;

   (e)Partition functions for Statistical Mechanics , with micro 
   partitions and Phase transitions, cond-mat/0411176;

   Non commutative quantum spacetime with topological
   vortex states and dark matter in the universe , gr-qc/0310136

[2]Charles Bennett,Cosmology from start to finish,Nature,
   p1126 to 1131, vol440,27 April,2006

[3]Varun Sahni,Dark Matter and Dark Energy,
    www.arxiv.org/astro-ph/0403324

[4]S.P.Misra , Introduction to supersymmetry and super gravity,
   Wiley Eastern(1992)

[5]C Rounanas, A.Maniero,D.V.Nanopoulus,K A Olive,eds (1984)
   Grand Unification with SUSY and cosmology, WSP.

[6]Tulsi Dass, Symmetries, Gauge fields, strings and fundamental
   interactions, Wiley Eastern(1993)

[7]K.A.Mitton,Physical manifestation of optical energy, the Casimir
   effect,  WSP(2001)

[8]J.Ellis,K.A.Olive,Y.Santoso,VC Spanos, Update on direct detection of 
   supersymmetric dark matter, Phys Rev Dv.71,095007(2005)

[9]Report of the task force on Cosmic microwave background research
  (2005)   <http://www.nsf.gov/mps/ast/tfcr.jsp>

[10]P.K Peebles,B.Ratra,The cosmological constant and dark energy,

    Rev.Mod.Phys.75,559-606(2003)

[11]P.S.Aspinwall,B.R.Greene,D.R.Morrison www.arxiv.org/hep-th/9311186
    -Spacetime topology change,physics of Calabi Yau moduli spaces

[12]P.Aspinwall, The breakdown of topology at small scales,
    www.arxiv.org/hep-th/0312188

[13]P.D Eath,Supersymmetric quantum cosmology,(1995),
   Cambridge university Press

[14]Carlo Rovelli, Quantum Gravity, Cambridge University Press(2004)

[15]John Peacock, Cosmological Physics, Cambridge university press(1999)

[16]L.X.Xu,H.Y.Liu,C.W.Zhang,5 dimensional cosmological models,
    and dark energy,215,v15,12 Feb 2006,
    International Journal of Modern PhysicsD
 
[17]N.Straumann Dark energy review www.arxiv.org gr-qc/0311083

[18]D.N.Spergel et al March 17(2006),Three Year Wilkinson Microwave
    Anisotropy Probe Observations: Implications for Cosmology,

    www.map.gsfc.nasa.gov/m- mm/pub- papers/threeyear.html

[19]N.D.Birrell,PCM Davies, Quantum fields in curved space,
    Cambridge university Press(1984)

[20]G.Gibbons,EPS Shellard,SJ Rankin,Eds, The future of Theoretical
    physics and cosmology, Cambridge university press (2003)

[21] V M Mostapaneko,NN Trunov,The Casimir effect and its 
    applications,WSP (2002)

[22] T.Padmanabhan,Physics Reports,(2003),380,235

\end{document}